\documentclass[twocolumn,showpacs,preprintnumbers,amsmath,amssymb]{revtex4}

\usepackage{graphicx}
\usepackage{dcolumn}
\usepackage{bm}
\usepackage{subfigure}

\begin{document}

\author{Alexandre E. Hartl}
 \altaffiliation[ ]{Department of Mechanical and Aerospace Engineering, North Carolina State University, Raleigh, NC 27695-7910.}

\author{Bruce N. Miller}
 \altaffiliation[ ]{Department of Physics and Astronomy, Texas Christian University, Forth Worth, TX 76129}
 \email{b.miller@tcu.edu}
 \homepage{http://personal.tcu.edu/ bmiller}

\author{Andre P. Mazzoleni}
 \altaffiliation[ ]{Department of Mechanical and Aerospace Engineering, North Carolina State University, Raleigh, NC 27695-7910.}

 \date{\today}

\title{Dynamics of an Inelastic Gravitational Billiard with Rotation}

\begin{abstract}

The seminal physical model for investigating formulations of nonlinear dynamics is the billiard. Gravitational billiards provide an experimentally accessible arena for their investigation. We present a mathematical model that captures the essential dynamics required for describing the motion of a realistic billiard for arbitrary boundaries, where we include rotational effects and additional forms of energy dissipation. Simulations of the model are applied to parabolic, wedge and hyperbolic billiards that are driven sinusoidally. The simulations demonstrate that the parabola has stable, periodic motion, while the wedge and hyperbola (at high driving frequencies) appear chaotic.  The hyperbola, at low driving frequencies, behaves similarly to the parabola; i.e., has regular motion.  Direct comparisons are made between the model's predictions and previously published experimental data.  The representation of the coefficient of restitution employed in the model resulted in good agreement with the experimental data for all boundary shapes investigated.   It is shown that the data can be successfully modeled with a simple set of parameters without an assumption of exotic energy dependence.

\end{abstract}

\pacs{05.10.-a;05.45.a;05.45.Ac;05.45.Pq}

\maketitle

\section{Introduction}

The seminal physical model for investigating formulations of nonlinear dynamics is the billiard. Traditionally this is a field free, classical, system where a particle experiences elastic collisions with a rigid boundary. Depending on the boundary shape, the ensuing motion can be stable or chaotic \cite{Bunomovich:1, Berry:1}. Quantum mechanical billiards have also been investigated and many of their essential features can be effectively represented with classical waves on membranes or cavities \cite{Stockman:1}. A limitation of the classical billiard is the difficulty in reproducing the system in the laboratory as a consequence of dissipation and the earth's ubiquitous gravitational field. In contrast, gravitational billiards provide an experimentally accessible arena for testing formulations of nonlinear dynamics.
\\
\indent
One and two-dimensional Hamiltonian versions of gravitational billiards have long provided easily visualized systems that exhibit a wide range of stable and chaotic behavior \cite{Barroso:1,Mehta:1,Miller:1,Korsch:1,Miller:2}. The system consists of a particle undergoing elastic collisions within a rigid boundary, where the particle follows a ballistic trajectory under the influence of a constant gravitational field between collisions. When the boundary is periodically driven, Fermi acceleration may result \cite{Lenz:1}, establishing a connection with cosmic ray physics and cosmology.  Billiards that bounce vertically on a level, oscillating surface have been used to model the impact process for numerous engineering applications including moving parts in machinery, impact dampers, fluid induced vibration in tubes and moored ships driven by steady waves \cite{Barroso:1,Luo:1,Holmes:1}.  Recent interest in dynamics has been focused on dissipative systems such as granular media. While inelasticity in these systems is usually represented by a collisional restitution coefficient, it has been observed that the inclusion of rotational friction induces qualitative changes in behavior \cite{Ben-naim:1} that cannot be explained by other means. Similarly, in billiard experiments \cite{Olafsen:1}, when friction is left out of the theoretical formulation, it appears that one is forced to make unphysical assumptions about the source of energy loss to approximately replicate the experimental data \cite{Gorski:1}.
\\
\indent
This paper considers the more realistic situation of an inelastic, rotating, gravitational billiard in which there are retarding forces due to air resistance and friction.  In this case the motion is not conservative, and the billiard is no longer a particle, but a sphere of finite size. We present a mathematical model that captures the relevant dynamics required for describing the motion of this ``real world" billiard for arbitrary boundaries.  The model is applied to parabolic, wedge and hyperbolic billiards that are driven sinusoidally.  Direct comparisons are made between the model's predictions and experimental data previously collected \cite{Olafsen:1}.  Although several studies have investigated the effect of variable elasticity in relation to the gravitational billiard, this study is the first to incorporate rotational effects and additional forms of energy dissipation.
\\
\indent
The ergodic properties of Hamiltonian gravitational billiards are well studied \cite{Miller:1,Korsch:1,Miller:2,Wojt:1}. It has been shown that the parabolic billiard is completely integrable having stable, periodic orbits \cite{Korsch:1}. Studies of the wedge billiard demonstrated that the billiard's behavior depends on the vertex angle, defined as $2\theta$ \cite{Miller:1}.  For 0 $<\theta<45^{o}$, the phase space contains coexisting stable and chaotic behavior. For $\theta = 45^{o}$ the motion is completely integrable, while for $\theta>45^{o}$ the motion is chaotic. Wojtkowski refers to this geometry as ``fat billiards" and has rigorously proven that they have a single, ergodic component \cite{Wojt:1}. These results have also been confirmed through experiments for an optical billiard with ultra cold atoms \cite{Milner:1}.  It has been demonstrated with numerical simulation that the motion of a hyperbolic billiard exhibits characteristics of the parabolic billiard for low energy, where the motion is near the origin, and for the wedge billiard at high energy, where the motion is mostly concentrated at its asymptotic limits \cite{Miller:2}.
\\
\indent
Feldt and Olafsen\cite{Olafsen:1} have experimentally studied a real inelastic billiard for a variety of boundaries.  One experiment consisted of a steel ball moving within a closed reflective aluminum boundary shaped either as a parabola, wedge or hyperbola.  The container was driven in the horizontal direction to compensate for energy losses resulting from collisions.  Imaging software determined the ball's position and velocity at the collision points.  The study results indicated regular motion for the parabola and chaotic motion for the wedge; the motion for the hyperbola was found to be frequency dependent, sharing characteristics of the parabola at low-driving frequencies and the wedge for higher-driving frequencies.
\\
\indent
In this work direct comparisons are made between simulations of the model system and the experimental data of Feldt and Olafsen. To date, G$\acute{o}$rski and Srokowski\cite{Gorski:1} are the only investigators known to have theoretically studied the experiments conducted by Feldt and Olafsen.  In their model they consider an inelastic, gravitational billiard for the case of no friction (or rotation) and no drag. In order to replicate the main features of the experiments, it was necessary to resort to a surprising, unconventional representation of the restitution coefficient energy dependence.
\\
\indent
This paper begins with a discussion on Kane's equations and the impact theory used for describing a collision between a billiard and a moving boundary.  The Appendix develops the equations governing this collision process in detail.  Sections III and IV describe the trajectory model for tracking the billiard's motion after each bounce and the procedure for detecting collisions.  Section V explains how the coefficients of restitution and friction are determined for numerical simulations.  Section VI presents simulations comparing the numerical results to previous experiments.  Conclusions are presented in Section VII.

\section{Kane's Equations and the Impact Theory}

There are several competing models for describing a collision of a billiard with a boundary, and each model has its own merit depending on the intended application.  First, there are analytical models that determine the billiard's velocity post collision in terms of the pre-impact velocities.  The formulation is based on Newton's law of motion and Coulomb's law of friction, and requires knowledge (a priori) of the coefficients of restitution and friction.  Accepted models in this category include works by Walton\cite{Walton:1} and Kane and Levinson\cite{Kane:1}.  Second, there are impact models based on the field of continuum mechanics, which consider collisions of elastic, viscoelastic or plastic objects.  Models in this group are based on the Hertzian contact theory and its offshoots\cite{Gugan:1}.
\\
\indent
In this study we employ a modified version of the impact theory set forth by Kane and Levinson for collisions between the billiard and boundary\cite{Kane:1}, where we account for the effects of a moving boundary.  The original theory provides a direct method for computing the billiard's generalized speeds post collision considering fixed boundaries.   The theory is based on Kane's equations which utilize partial velocities and generalized forces for deriving equations of motion.  The equations are also known as Lagrange's form of D'Alembert's principle, and references \cite{Kane:1} and \cite{Baruh:1} provide a thorough treatment on the subject.
\\
\indent
Kane and Levinson make the following assumptions in their model: first, the contact area between the objects is a single point through which all forces are exerted.  Second, the total collision impulse is represented by the integral of the forces over the entire collision time.  Third, the coefficients of restitution, static friction and kinetic friction are constants to be determined experimentally. The theory initially assumes no slipping at the contact point between the sphere and boundary.  A set of values for the generalized speeds are generated, and are valid if and only if the no-slip condition is satisfied.  If the no-slip condition is violated, then a new set of values for the generalized speeds are developed under the assumption of slipping.  See reference \cite{Kane:1} for a detailed derivation of the theory.
\\
\indent
We extend Kane and Levinson's impact theory to include collisions on moving boundaries.  For completeness, we list the equations of motion for this system in the Appendix.  Next, we introduce a trajectory model that tracks the billiard's motion between bounces, taking into account dissipative and aerodynamic forces.  By employing the impact and trajectory models, we then construct an efficient set of algorithms describing the billiard's motion in order to perform numerical simulations.

\section{Trajectory Model and the Reinitialization of the Generalized Speeds}

Between collisions we make use of a trajectory model that numerically tracks the billiard's motion after each bounce.  The model is used to reinitialize the generalized speeds at the point of initial contact with the boundary.  As a starting point, we define the equations of motion that governs the billiard's movement while airborne.
\\
\indent
For a billiard (or sphere) moving through air, its motion is affected by gravity, air resistance (drag) and additional aerodynamic forces due to its spinning motion (the Magnus effect).  In this study the Magnus Force is neglected since its overall effect is small.  The force of gravity acting on the billiard is defined as
\begin{equation}
\label{106:functionc}
\mathbf{F}_{G}=-mg\mathbf{n}_{2}
\end {equation}
where $m$ is the billiard's mass and $g$ is the acceleration due to gravity.  Refer to the Appendix for a definition of the coordinate system.
\\
\indent
The drag force exerted on the billiard is a function of the billiard's velocity, and acts in the direction opposite to its path.  At low velocities the drag force is linearly proportional to the billiard's speed, but shows a quadratic dependence on speed at higher velocities.  Generally, the drag force acting on a body is determined by experimental measurements, and is often approximated by the equation
\begin{equation}
\label{1102:u}
\mathbf{F}_{D}=-\mathbf{v}\left(c_{1}+c_{2}\left|\mathbf{v}\right|\right)
\end{equation}
where $c_{1}$ and $c_{2}$ are constants that are dependent on the size and shape of the object \cite{Fowles:1}.
Traditionally, $c_{1}$ and $c_{2}$ are expressed as
\begin{eqnarray}
\label{1104:u}
&& c_{1} = 6\pi\eta b \nonumber  \nonumber \\
&& c_{2} = \frac{1}{2}\rho A C_{D}
\end{eqnarray}
where $\eta$ is the dynamic viscosity of air, $b$ is the object's radius, $\rho$ is the atmospheric density, $A$ is the object's cross-sectional area and $C_{D}$ is the drag coefficient.  In Equation~(\ref{1104:u}), $c_{1}$ is the coefficient of the familiar Stokes drag force, which is valid at low Reynolds number.  The drag coefficient, $C_{D}$, is a function of the Reynolds number.  Experimentally, $c_{1}$ and $c_{2}$ have been measured directly, giving the correct dependence on the object's diameter.  For spheres in air, approximate values for $c_{1}$ and $c_{2}$ in SI units are
\begin{eqnarray}
\label{1103:u}
&& c_{1} = 1.55 \times 10^{-4}D  \nonumber \\
&& c_{2} = 0.22D^{2}
\end{eqnarray}
where $D$ is the sphere's diameter in meters \cite{Fowles:1}.
In this study constants $c_{1}$ and $c_{2}$ are specified by Equation~(\ref{1103:u}), but note that the constants given by Equation~(\ref{1104:u}) yield similar results.  The ratio of the quadratic term to the linear term of the drag force, i.e.,
\begin{equation}
\label{1006:functionc}
\frac{c_{2}\mathbf{v}\left|\mathbf{v}\right|}{c_{1}\mathbf{v}} = 1.4 \times 10^{3} \left|\mathbf{v}\right| D
\end {equation}
determines which type of drag is more significant.  If the ratio is greater than one, the quadratic term is dominant.  If the ratio is below 1, the linear term is dominant.  If the ratio is around one, however, both terms must be taken into account.
\\
\indent
Subsequently, the equations of motion for a billiard traveling through air is
\begin{equation}
\label{107:functionc}
\mathbf{F}=\mathbf{F}_{G}+\mathbf{F}_{D}
\end {equation}
The billiard and boundary are simulated using a time-driven procedure, where the system is advanced in time until a collision is detected \cite{Hartl:1,Hartl:2}.  Between boundary encounters the trajectory equations denoted by equation~(\ref{107:functionc}) consist of second-order, nonlinear, coupled differential equations which are solved numerically by using a fourth-order Runge-Kutta method.

\section{Collision Detection Method}

\begin{figure}[htb]
  \begin{center}
    \leavevmode
      \includegraphics[width=3.55in]{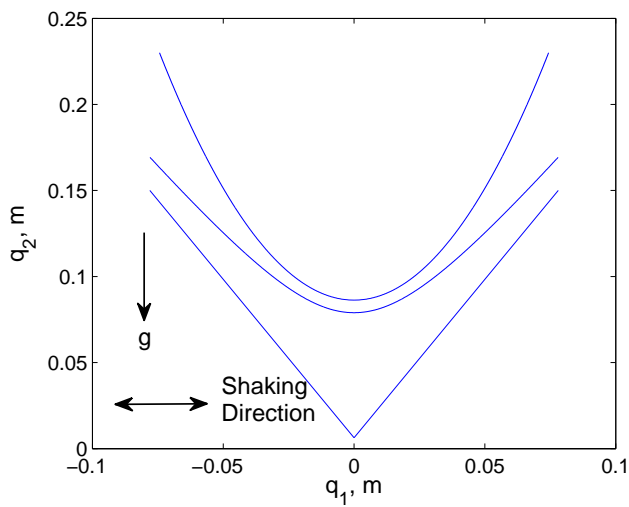}
    \caption{A schematic of the boundaries used in the simulations.}
    \label{fig:schematic}
  \end{center}
\end{figure}

In this section we outline a general procedure for detecting collisions between the billiard and a boundary of arbitrary shape.  The procedure locates the minimum distance between the objects at each time step, and compares that distance to a specified tolerance, which for our case is the billiard's radius $b$.  If the distance is less than or equal to the tolerance, then a collision is reported.  Otherwise, it is concluded that no collision has occurred.
\\
\indent
A detailed description of the procedure now follows:  first, write the square of the distance formula $L^{2}$ between the billiard and boundary in terms of the billiard's geometric center and the mathematical formula that describes the boundary's shape.  Second, use the boundary formula to eliminate all but one of the variables, thereby expressing $L^{2}$ as a function of a single variable.  Third, minimize $L^{2}$ by taking its derivative and setting it equal to zero.  Fourth, solve for the roots of the resulting equation, where valid solutions are restricted to the set of real numbers.  (Depending on the boundary, the roots may be determined by analytical or numerical methods).  Fifth, locate the minimum distance between the objects by substituting the roots into $L^{2}$.  (If $L^{2} \leq b^2$, then the billiard is impacting the boundary.  Otherwise, the objects are not colliding).
\\
\indent
If a collision is detected, the collision time and collision location are approximated by interpolation methods.  The procedure for finding the collision time is based on a paper by Baraff \cite{Baraff:1}.   It searches for a configuration where the penetration depth between the objects is sufficiently close to zero.  As a consequence the determination of the collision time is transformed into a root-finding problem, where the system's state at the collision time is approximated by interpolating the derivatives computed by the Runge-Kutta method.
\\
\indent
In the following we consider driven parabolic, wedge and hyperbolic boundaries defined mathematically (in the laboratory frame), respectively, as:
\begin{eqnarray}
\label{1033:u}
&& q_{2} = f(q_{1}) = a\left(q_{1} - \Delta q_{1}\right)^{2} + c   \\
&& q_{2} = f(q_{1}) = b\left|q_{1} - \Delta q_{1}\right| + c  \\
&& q_{2} = f(q_{1}) = \sqrt{\alpha \left(1 + \beta \left(q_{1} - \Delta q_{1}\right)^{2}\right)} - \delta \
\end{eqnarray}
where $a = 0.26 cm^{-1}$, $b = 1.85$, $c = 0.63 cm$, $\alpha = 40.3 cm^{2}$, $\beta = 0.08 cm^{-2}$ and $\delta = 4.45 cm$.  These are the values used in the experiments of Feldt and Olafsen.  The boundaries oscillate horizontally and their position at time $t$ is defined by
\begin{equation}
\label{1034:functionc}
\Delta q_{1}\left(t\right) = A sin \omega t
\end {equation}
where $A$ is the amplitude and $\omega = 2\pi f$ is the oscillation angular frequency in rad/sec and $f$ is the oscillation frequency in hertz.  Experimentally, note that the boundaries are sealed off by an aluminum top and thin pieces of Plexiglas on the sides, rendering the system effectively two-dimensional.  Figure~\ref{fig:schematic}  shows the boundary shapes and their orientation with respect to gravity and the driving direction. Note the boundaries are offset and their horizontal tops are omitted for clarity.
 The billiard's position is tracked in time by following its geometric center, where its position is defined by
\begin{equation}
\label{1011:eq}
\mathbf{q}=q_{1}\mathbf{n_{1}}+q_{2}\mathbf{n_{2}}+q_{3}\mathbf{n_{3}}
\end {equation}
For the numerical simulations presented in this study, $q_{3} = 0$.  In reality, however, $q_{3} \neq 0$ because of slight chattering in the $\mathbf{n_{3}}$ direction.  Private communications with an author of \cite{Olafsen:1} reveal that the system noise induced by this effect is small.  The actual billiard considered in the simulations and used in the experiments is a $3.1$ mm diameter steel chrome ball weighing approximately $0.13028$ g.
\\
\indent
Application of the above procedure results in solving a second-order algebraic equation for the wedge, a third-order algebraic equation for the parabola and a fourth-order algebraic equation for the hyperbola.  We determine the roots of the equations by means of the Newton-Raphson method.

\section{The Coefficients of Restitution and Friction}
\begin{figure}[!htp]
  \begin{center}
    \leavevmode
      \includegraphics[width=3.55in]{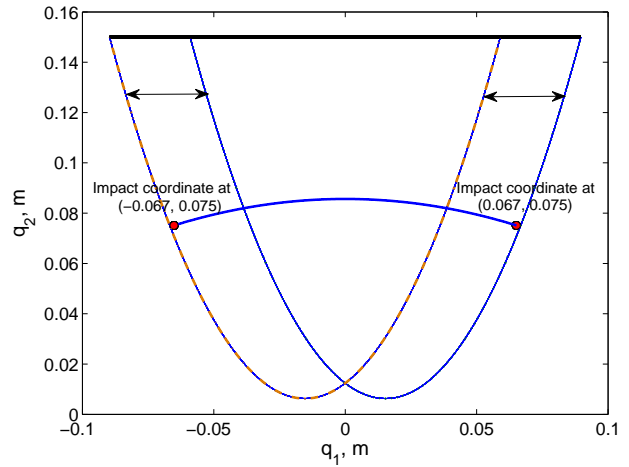}
    \caption{The trajectory of the stable period-one orbit and the location of the parabolic boundary at the impact points.}
    \label{fig:parab_1}
  \end{center}
\end{figure}

\begin{figure}[!htp]
  \begin{center}
    \leavevmode
      \includegraphics[width=3.55in]{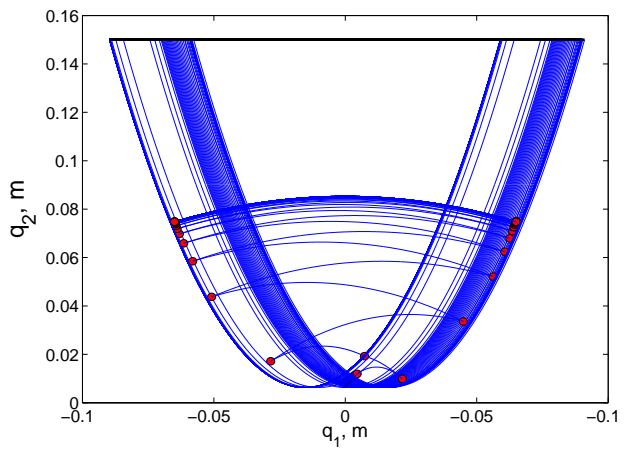}
    \caption{The evolution of the billiard's trajectory for achieving a stable period-one orbit.}
    \label{fig:para_e2}
  \end{center}
\end{figure}

Collisions with the boundary result in energy losses stemming from the restitution in the normal direction and friction in the transverse direction.  Feldt and Olafsen \cite{Olafsen:1} suggest a coefficient of restitution of 0.9 between the steel billiard and the aluminum boundary, noting that its value is velocity dependent.  However, as we see in the following, the coupling between the normal and tangential contact forces during the impact reduces this coefficient considerably. Further, it has been observed that the coefficient of restitution depends on the incident angle at impact \cite{Louge:1}.  Friction, due to confining walls, also plays a vital role in the experiments of Feldt and Olafsen.  Studies on granular media show the importance of including frictional effects between particles and their containers since the particles' velocity distributions are affected by these interactions\cite{Swinney:1,Menon:1}.  Kane and Levinson's impact theory remains practical when it is supplemented by experimental measurements capturing the coefficients of restitution and friction.  Toward this end the parabolic billiard is used as a standard test case for establishing the coefficient of restitution since experiments have shown that it exhibits stable, period-one orbits.  The experiment described in \cite{Olafsen:1} resulted in an orbit height of approximately $7.5$ cm; an apparent value for the coefficient of restitution is estimated by matching the orbit height of the simulation to the experiment.  First, note that for steel on aluminum, experiments reveal that the coefficient of static and kinetic friction is approximately $0.61$ and $0.47$ \cite{eng:1}, respectively.  If the numerical model applies a coefficient of restitution of $e = 0.393$ along with the friction coefficients specified above, then the simulation approximately replicates the experiment; as a result we apply this $e$ value for all boundary shapes considered in this paper.  Note that if the effects of air resistance are omitted, then the coefficient of restitution drops slightly to $e = 0.392$.  This result is not surprising considering the smallness of the billiard and its relatively short time of flight between bounces.  Figure~\ref{fig:parab_1} shows the trajectory of the stable orbit and the location of the parabolic boundary at the impact points, while Figure~\ref{fig:para_e2}  reveals the evolution of the billiard's trajectory for entering a period-one orbit after starting at the origin.   If sufficient energy is supplied to the system, the billiard's trajectory eventually mode-locks into a stable period-one orbit. The orbit moves up or down the parabola as the driving frequency is increased or decreased, respectively.  If insufficient energy is given to the system, then the parabola will explore multiple trajectories.
\\
\indent
If one ignores friction, the coefficient of restitution required to match the orbit height from the experiment drops significantly to $e = 0.246$.  If air resistance is also neglected, then the coefficient of restitution drops to $e = 0.245$.  Table~\ref{Table_1} summarizes the different values of the coefficient of restitution considering several dynamical effects.  For the amplitude and frequencies considered in this paper, the coefficient of restitution is dominated by frictional effects, while air drag plays a negligible role.  This demonstrates a greater need to understand the relationships between the friction and restitution coefficients.
\begin{table}[!htb]
\begin{center}
\caption{The coefficient of restitution required to reach a stable, period-1 orbit for a parabolic billiard considering various dynamical effects. X indicates the effect is included; $---$ indicates the effect is omitted.}
\begin{tabular}{c c c} \hline \hline
\textbf{Friction}         &     \textbf{Drag}          &     \textbf{Coefficient of Restitution} \\ \hline
  $---$ & $---$ & 0.245   \\
  $---$ & X & 0.246  \\
  X & $---$ & 0.392   \\
  X & X & 0.393  \\ \hline
\end{tabular}
\label{Table_1}
\end{center}
\end{table}
\\
\indent
G$\acute{o}$rski and Srokowski suggest a coefficient of restitution of $e = 0.43$ for the no friction, no drag case.  However, their approach for determining the billiard's velocity change post collision is unconventional since they apply the coefficient of restitution to the complete velocity, instead of only to the normal component of velocity.  Without friction, the parallel component of momentum must be conserved.  To confirm their assumption, examine Equation 1 of their paper, which reproduced here is
\begin{equation}
\label{400:eq}
\mathbf{v}^{C}_{1} = r\left(\mathbf{v}^{C}_{0} - 2\mathbf{u}\left(\mathbf{v}^{C}_{0}\cdot \mathbf{u}\right)\right)
\end {equation}
where $\mathbf{v}^{C}_{0}$ and $\mathbf{v}^{C}_{1}$ are the particle's velocity before and after a collision, respectively, $\mathbf{u}$ is the velocity normal to the boundary and $r$ is the coefficient of restitution \cite{Gorski:1}.  Now, suppose $\mathbf{v}$ is the velocity tangent to the boundary.  Taking the scalar product of Equation~(\ref{400:eq}) with respect to $\mathbf{u}$ and $\mathbf{v}$ results in the following expressions:
\begin{eqnarray}
\label{401:eq}
& \mathbf{v}^{C}_{1} \cdot \mathbf{u}& = r\left[\mathbf{v}^{C}_{0} \cdot \mathbf{u} -2\mathbf{u} \cdot \mathbf{u} \left(\mathbf{v}^{C}_{0} \cdot \mathbf{u}\right)\right]   \nonumber\\
&& =r\left[\mathbf{v}^{C}_{0} \cdot \mathbf{u}\right]\left[1-2\right]\nonumber\\
&& =-r\left[\mathbf{v}^{C}_{0} \cdot \mathbf{u}\right]
\end{eqnarray}
\begin{eqnarray}
\label{402:eq}
& \mathbf{v}^{C}_{1} \cdot \mathbf{v}& = r\left[\mathbf{v}^{C}_{0} \cdot \mathbf{v} -2\mathbf{u} \cdot \mathbf{v} \left(\mathbf{v}^{C}_{0} \cdot \mathbf{v}\right)\right]   \nonumber\\
&& =r\left[\mathbf{v}^{C}_{0} \cdot \mathbf{v}\right]
\end{eqnarray}
By examining Equations~(\ref{401:eq}) and~(\ref{402:eq}), it is clear that the coefficient of restitution is incorrectly applied to both the normal and tangential components of the velocity.  As a consequence comparison of their results to the experiments performed by Feldt and Olafsen remain ambiguous.
\\
\indent
In granular media simulations, a method of preventing inelastic collapse of particles is to set the coefficient of restitution to its elastic limit of $1$ if collisions occur too frequently \cite{Luding:1}.  Additionally, studies have demonstrated that the coefficient of restitution approaches a value of $1$ as the normal component of the impact velocity approaches $0$.  As a consequence we apply a coefficient of restitution of $1$ in our simulations if the relative velocity (between the billiard and boundary) in the normal direction is sufficiently small that it results in inelastic collapse, where a ``nearly infinite'' number of collisions occur in a finite time \cite{Luding:1}.  In practice, this assumption is only applied at the start of simulations (to initiate motion) and for brief instances of time during the simulation (as explained above).

\section{Simulations}

\begin{figure}[htb]
\begin{center}
\subfigure{\includegraphics[scale=0.4]{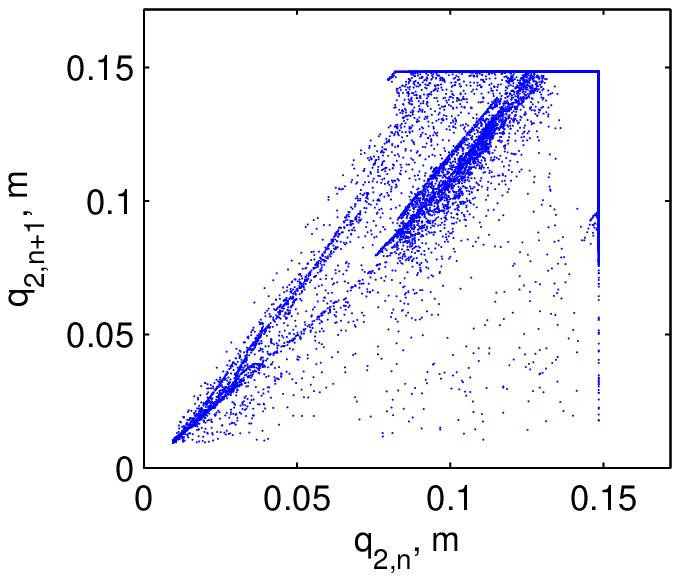}}
\subfigure{\includegraphics[scale=0.4]{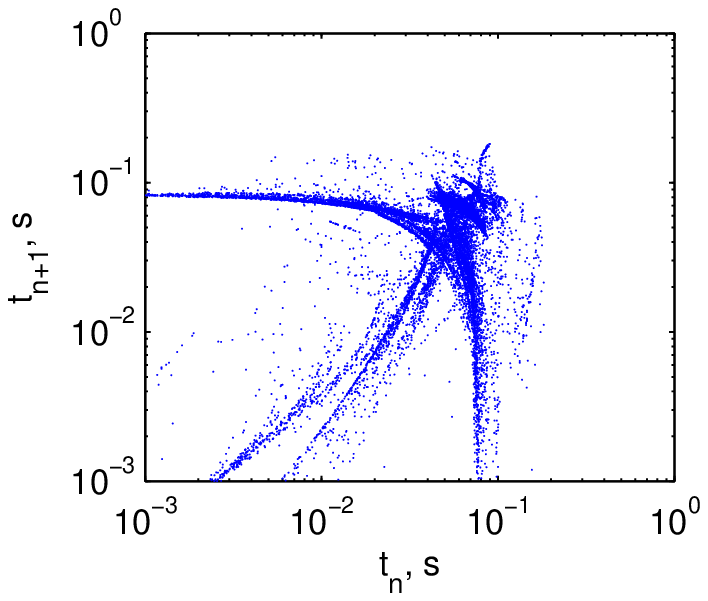}}\\
\subfigure{\includegraphics[scale=0.4]{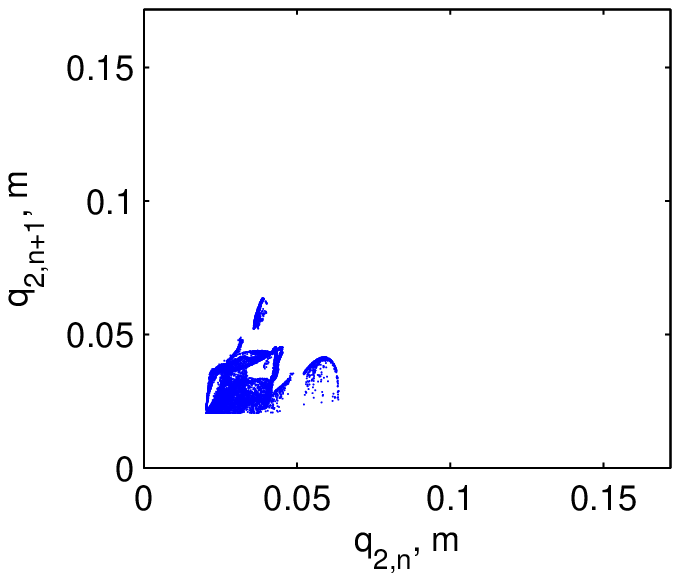}}
\subfigure{\includegraphics[scale=0.4]{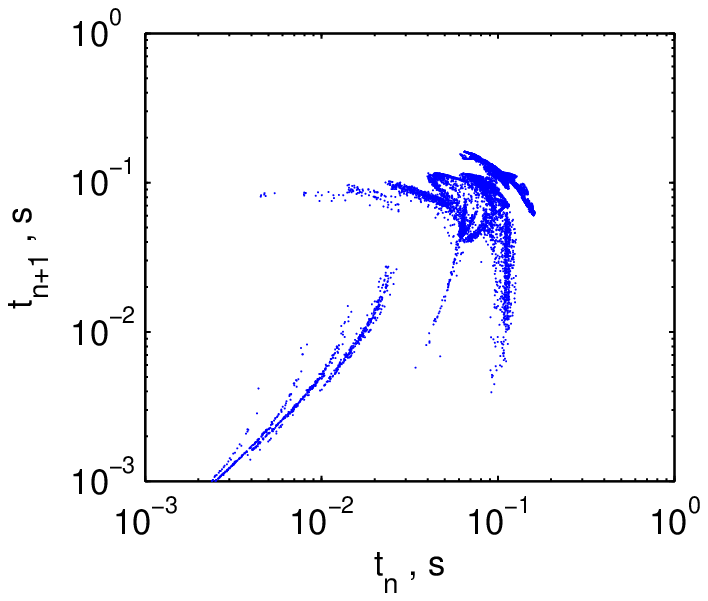}}\\
\subfigure{\includegraphics[scale=0.4]{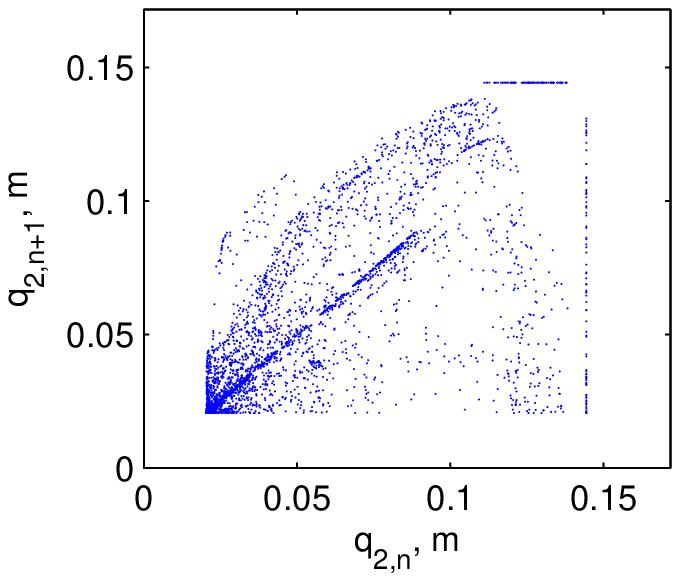}}
\subfigure{\includegraphics[scale=0.4]{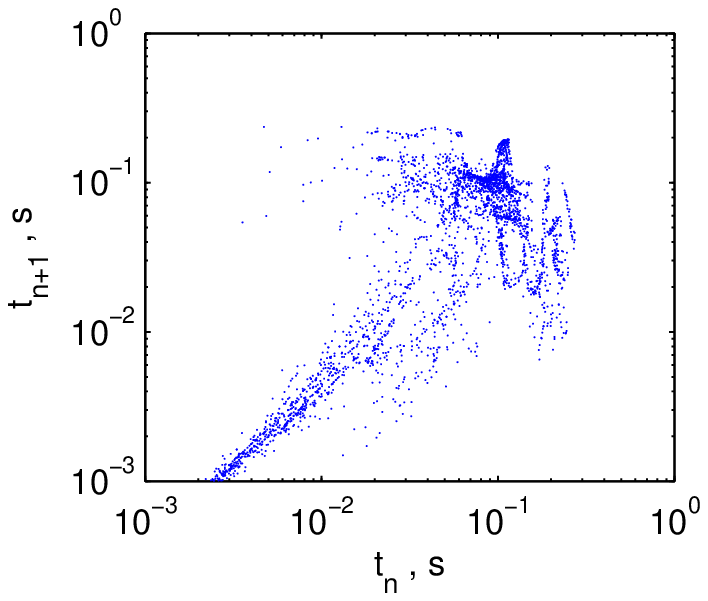}}\\
\subfigure{\includegraphics[scale=0.4]{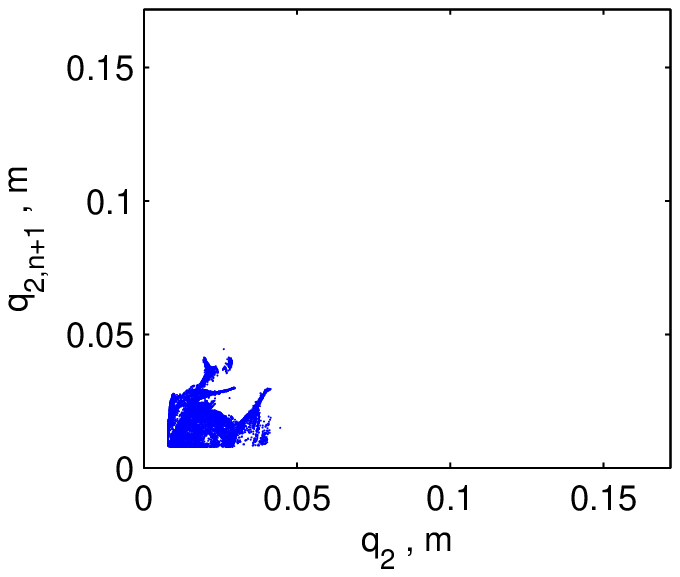}}
\subfigure{\includegraphics[scale=0.4]{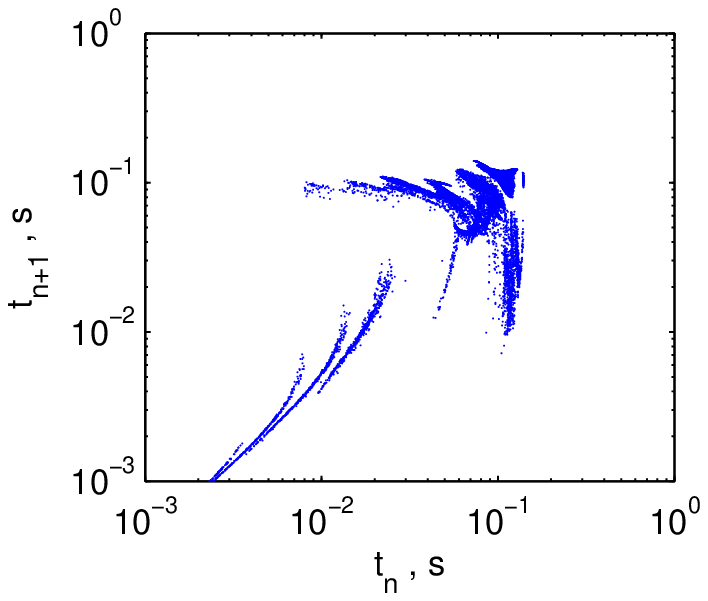}}
\end{center}
\caption{The spatial and temporal mappings of the collision heights (left column) and times of flight (right column).  From top to bottom: the wedge at 6.6 Hz, the hyperbola at 4.5 Hz, the hyperbola at 5.8 Hz, the parabola at 4.5 Hz.}
\label{fig:map_1}
\end{figure}

\begin{figure}[htb]
\begin{center}
\leavevmode
\includegraphics[width=1.75in]{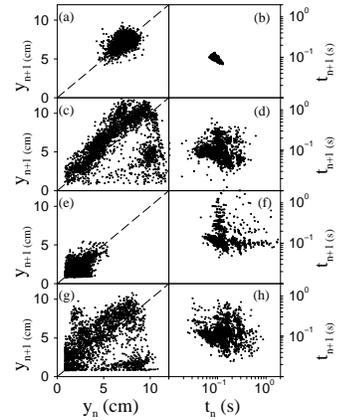}
\end{center}
\caption{The spatial and temporal mappings of the collision heights (left column) and times of flight (right column) for the experimental data of Feldt and Olafsen \cite{Olafsen:1}.  From top to bottom: the parabola at 5.4 Hz, the wedge at 6.6 Hz, the hyperbola at 4.5 Hz the hyperbola at 5.8 Hz.}
\label{fig:Olafsen_fig}
\end{figure}

\begin{figure}[htb]
\begin{center}
\subfigure{\includegraphics[scale=0.4]{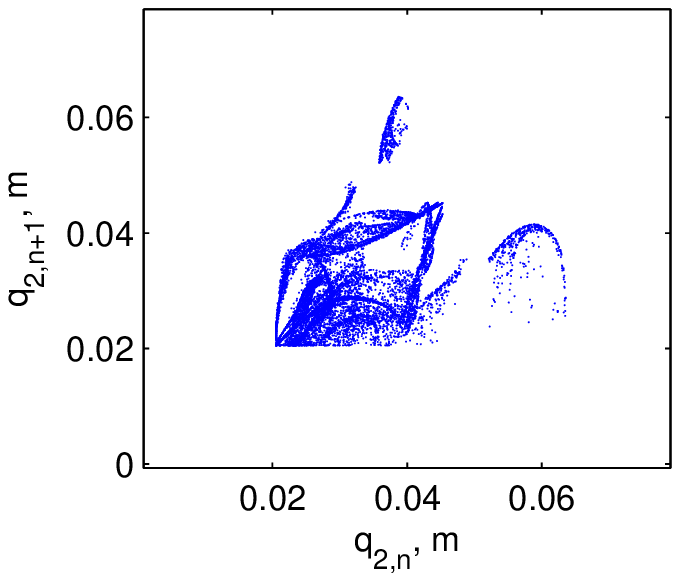}}
\subfigure{\includegraphics[scale=0.4]{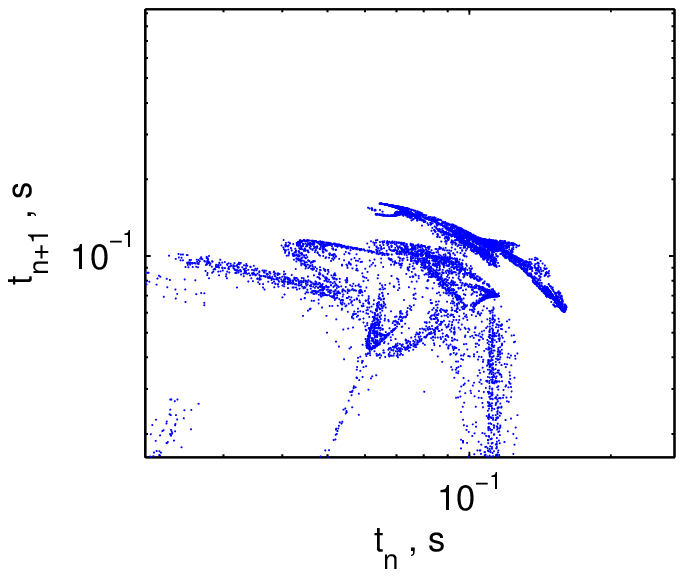}}\\
\subfigure{\includegraphics[scale=0.4]{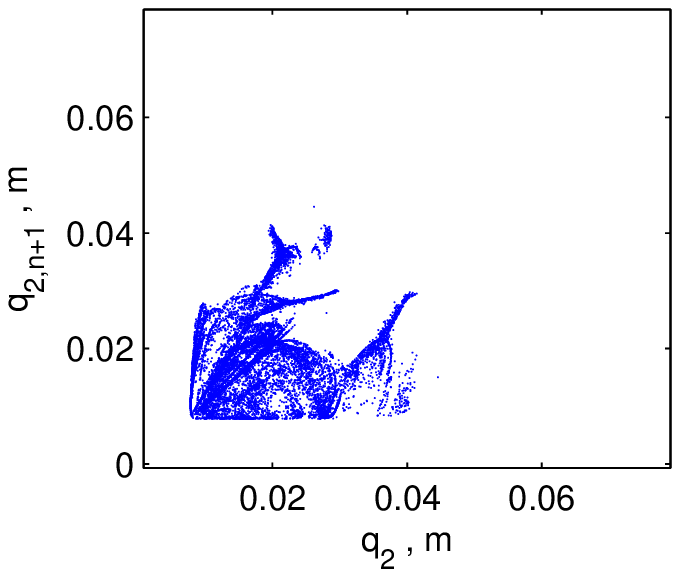}}
\subfigure{\includegraphics[scale=0.4]{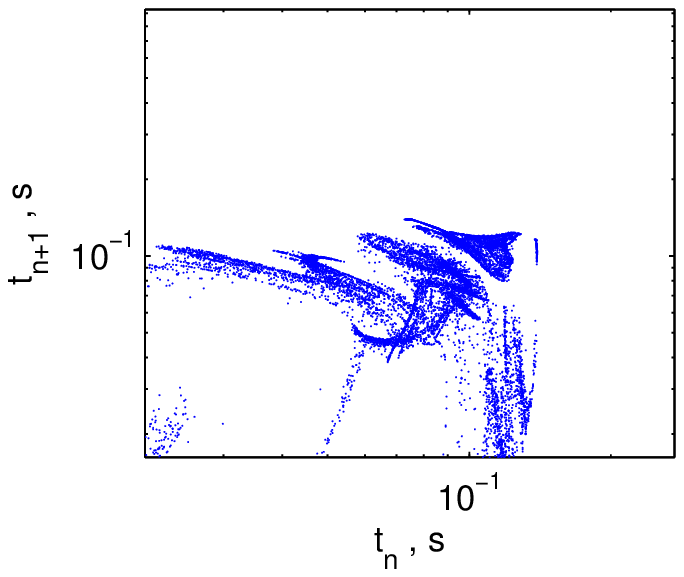}}
\end{center}
\caption{A close-up of the spatial and temporal mappings of the collision heights (left column) and times of flight (right column) for the hyperbola (top row) and parabola (bottom row) at 4.5 Hz.}
\label{fig:map_2}
\end{figure}

\begin{figure}[htb]
\begin{center}
\subfigure{\includegraphics[scale=0.4]{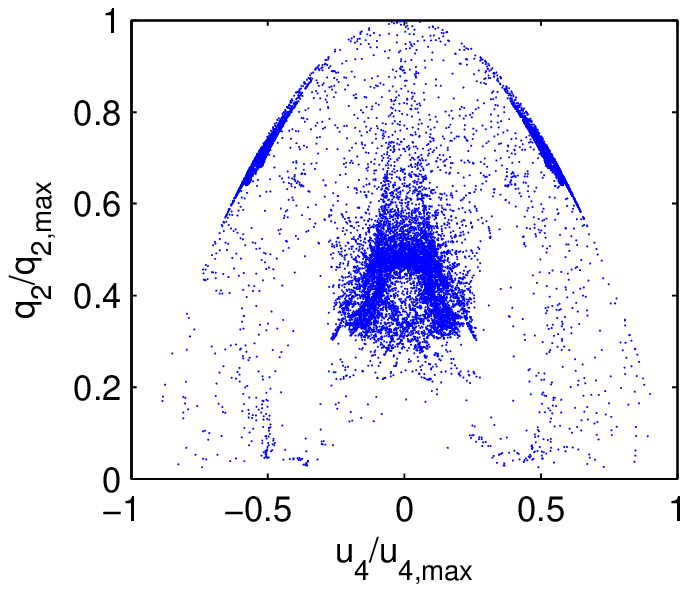}}
\subfigure{\includegraphics[scale=0.4]{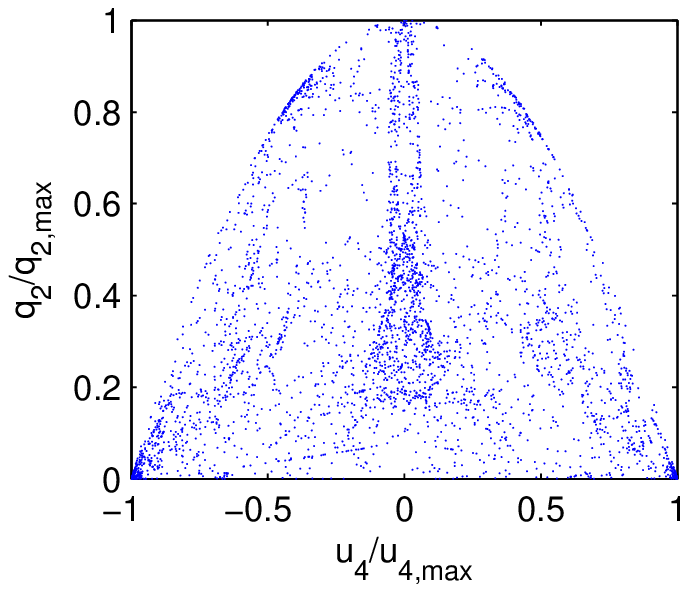}}\\
\subfigure{\includegraphics[scale=0.4]{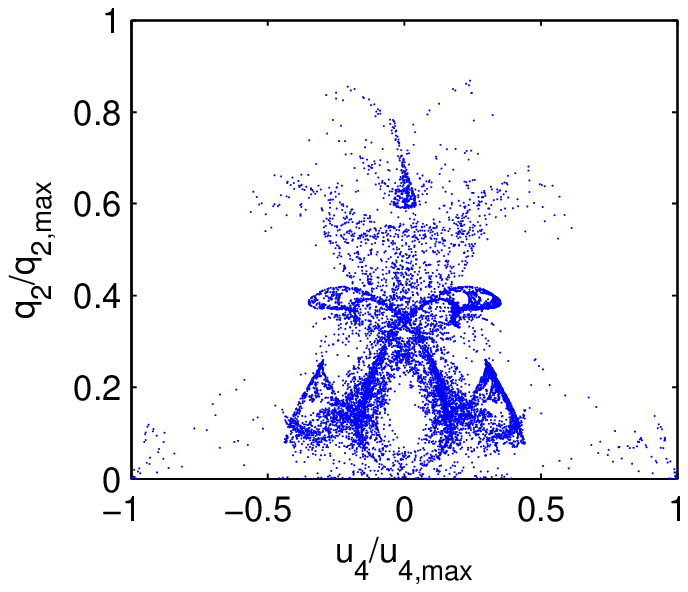}}
\subfigure{\includegraphics[scale=0.4]{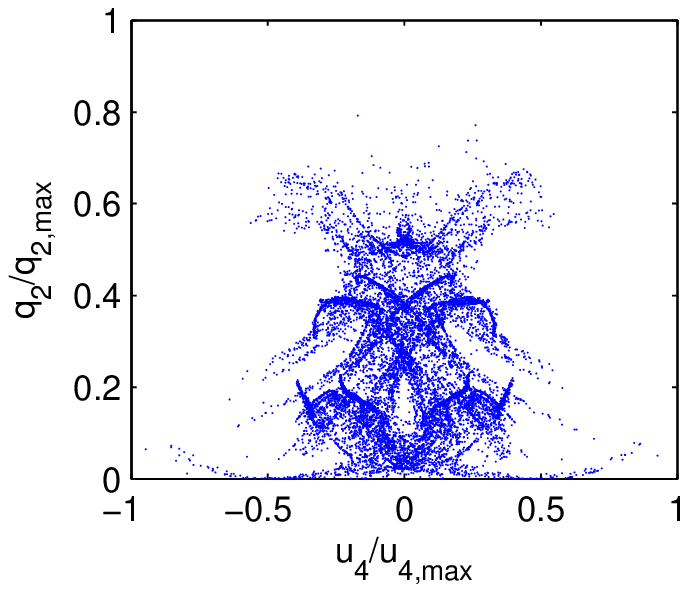}}
\end{center}
\caption{The normalized collision heights $q_{2}/q_{2,max}$ versus the normalized tangential velocity $u_{4}/u_{4,max}$.  Top row: the wedge at 6.6 Hz, the hyperbola at 5.8 Hz.  Bottom row: the hyperbola at 4.5 Hz, the parabola at 4.5 Hz.}
\label{fig:map_3}
\end{figure}

\begin{figure}[!htp]
  \begin{center}
    \leavevmode
      \includegraphics[width=3.55in]{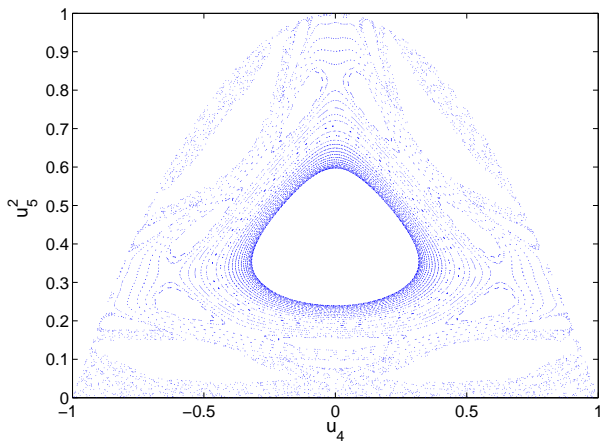}
    \caption{Surface of section for the Hamiltonian gravitational billiard for a wedge half-angle equal to $28.5^{o}$.}
    \label{fig:wedge_86}
  \end{center}
\end{figure}

\begin{figure}[htb]
\begin{center}
\subfigure{\includegraphics[scale=0.4]{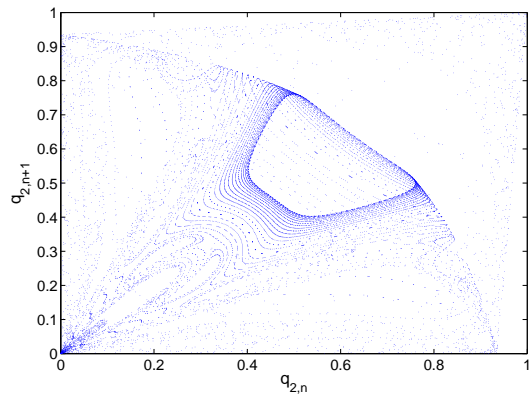}}
\subfigure{\includegraphics[scale=0.4]{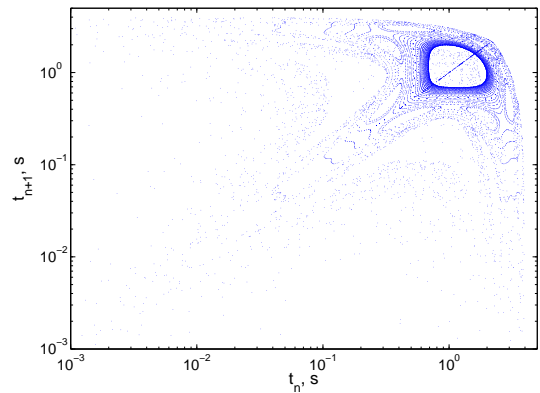}}\\
\subfigure{\includegraphics[scale=0.4]{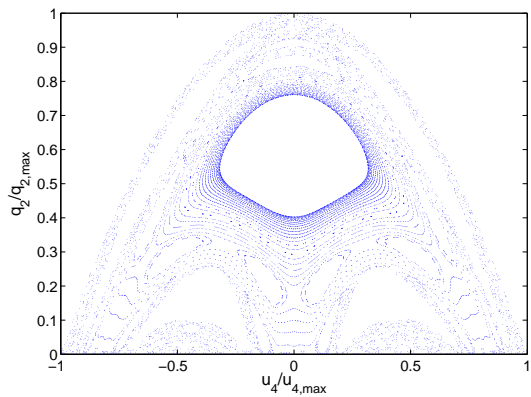}}
\end{center}
\caption{Additional mappings of the Hamiltonian gravitational billiard for a wedge half-angle equal to $28.5^{o}$.  The spatial and temporal mappings of the collision heights (top row), times of flight (middle row) and the normalized collision heights  versus the normalized tangential velocity (bottom row) are shown.}
\label{fig:map_4}
\end{figure}

Each simulation tracks a single trajectory consisting of 25,000 billiard-boundary collisions.  The billiard is initially at rest, but is quickly propelled into the air by the energy transmitted from the boundary to the billiard.  The collision height $q_{2}$ and time $t$ between consecutive bounces are extracted from the simulations and are shown in Figure~\ref{fig:map_1} for multiple boundaries driven at varying frequencies. For comparison, the experimental results from Feldt and Olafsen \cite{Olafsen:1} are given in Figure~\ref{fig:Olafsen_fig}. The successive mappings of the collision heights and times of flight show good resemblance to the experimental data even though a constant coefficient of restitution is used. In reality, the coefficient of restitution is velocity dependent and
 its representation is affected by the coupling between the normal and tangential contact forces\cite{Brilliantov:1}. The numerical model, however, detects additional collisions (i.e., collisions that occur in rapid succession) not reported by the experiments.  This is due to the lower resolution of the experimental data.  As a result the numerical model observes more collisions at both the small and large values of the height $q_{2}$; thus the time mappings have longer time tails associated with shorter times of flight for these collisions.
\\
\indent
The plot of the wedge driven at 6.6 Hz reveals that the motion appears chaotic as suggested by the experiments. The billiard is continuously driven to the top of the wedge, with most of the collision points laying above the $q_{2,n} = q_{2,n+1}$ line in the return map.  The time map also shows indications of chaotic behavior and a similar concentration of points to the experimental data. The hyperbola driven at 5.8 Hz resembles the unstable behavior of the wedge in both position and time, and shows a likeness to the experimental results.  For lower driving frequencies, the hyperbolic billiard is well approximated by the parabolic billiard as seen in Figure~\ref{fig:map_1}.  At 4.5 Hz, the billiard's motion is confined to the regions near the hyperbola's vertex, and shows semblance of a regular pattern not noted in the experimental data due to possible smearing of the data.  Patterns are also detected in the temporal mapping of the hyperbola at this driving frequency.  Figure~\ref{fig:map_2} displays these patterns in both the spatial and temporal mappings for the hyperbolic and parabolic billiards.  Note that the parabolic billiard driven at 4.5 Hz is not studied experimentally, but is used to demonstrate that the hyperbolic billiard behaves similarly to the parabolic billiard at low driving frequencies.
\\
\indent
Following the experiments the phase space is further investigated by plotting the normalized collision height $q_{2}/q_{2,max}$ versus the normalized tangential velocity $u_{4}/u_{4,max}$ post collision. The quantities $q_{2,max}$ and $u_{4,max}$ are defined as the maximum energy values that the billiard can possess at the collision height $q_{2}$ if all the energy were potential or kinetic, respectively.  A completely stable period-one orbit for a perfectly elastic billiard is characterized by having zero tangential velocity assuming collisions with a symmetric boundary.  For the parabolic billiard driven at 5.4 Hz, the numerical model predicts a small value for the normalized tangential velocity when the billiard achieves a stable period-one orbit; the mapping is a single point that has the value $\left(q_{2}/q_{2,max},u_{4}/u_{4,max}\right)$ =  $\left(0.376, \pm 0.0372\right)$.  The experimental data, however, shows that the normalized tangential velocity is a thin band about zero, where the range in velocity and height is caused by the noise in the system and small variations in the coefficient of restitution.  Figure~\ref{fig:map_3} shows the results for the remaining boundaries. For the case of the wedge, the billiard explores much of the phase space; the hyperbolic billiard driven at the higher frequency exhibits similar behavior to the wedge, but examines even more of the phase space.  For both shapes there are regions that have concentrations of points in the phase space not indicated by the experiments.  For the hyperbolic billiard driven at the lower frequency, regular patters develop, which are similar in appearance to the parabolic billiard driven at the same frequency.  The simulations of this paper for the hyperbolic billiard driven at 4.5 Hz is comparable to the results reported by G$\acute{o}$rski and Srokowski; the similarity exists because for low driving frequencies (or for low energy systems) the effects of their restitution assumption are mitigated.
\\
\indent
Note that the plots in Figure~\ref{fig:map_3} represents a significant deviation from the results indicated by the experiments.  The difference is qualitative and is not explained by the extra collisions reported by the numerical simulations.  Potentially, the source of the difference may lay in the finite resolution of the imaging software, resulting in measurement uncertainty, where the normalized collision height and normalized tangential velocity are calculated quantities that depend on the accurate resolution of the billiard's position, velocity and velocity components at the collision points.
\\
\indent
The experiments of Feldt and Olafsen were motivated by the earlier work on the wedge billiard \cite{Miller:1}. Of course, that system is Hamiltonian so there is no dissipation or friction. Moreover, there is no ceiling or upper boundary, and the system is not driven. With these caveats in mind, it is instructive to compare the predictions of the original model with the actual experiments. These are displayed in Figures~\ref{fig:wedge_86} and~\ref{fig:map_4} for a wedge with a half angle of $28.5^{o}$, exactly corresponding to the Feldt and Olafsen experiment. In Figure~\ref{fig:wedge_86} we show a Poincare surface where the square of the normal velocity is plotted vs the tangential velocity after each boundary collision. As  shown in the earlier work \cite{Miller:1}, this choice generates an area preserving map. The figure incorporates a number of distinct trajectories, all with a common energy. Surrounding a large, stable island associated with the period-one fixed point, we see a family of nested KAM tori all associated with the same fixed point. Surrounding this family are additional stability islands identified with different stable periodic points, as well as a space-filling chaotic orbit.
\\
\indent
In Figure~\ref{fig:map_4} we display the same data plotted with the alternative coordinate pairs employed in the Feldt and Olafsen experiment. Clearly there is greater structural detail in the Hamiltonian version than in the experiment (Figure ~\ref{fig:Olafsen_fig}) and simulations (Figure~\ref{fig:map_1}). This is not surprising because, in the driven system, there is a distribution of energies. Moreover, the dominant role of the fixed point is apparent in the former. In the plots of $q_{2,n+1} vs. \:q_{2,n}$ for the simulations (Figure~\ref{fig:map_1}) the role of upper-boundary collisions is apparent. These are not visible in the published experimental work (Figure~\ref{fig:Olafsen_fig}) because the scale height of the plots is too small.  It is intriguing that we see qualitatively similar behavior in the $t_{n+1} vs \:t_n $ plots for all three systems. Since the set of coordinate, $q_{2}/q_{2,mav} vs\: u_{2}/u_{2,max}$ in the final picture closely corresponds to Birkhoff coordinates, the similarity with Figure~\ref{fig:Olafsen_fig} is not surprising. In future work it would be interesting to investigate the effect of the upper boundary on a Hamiltonian wedge.

\section{Conclusions}

There are open questions concerning how best to model impacts between systems of solid objects, such as granular media. Examining the ergodic properties of a gravitational billiard provides an experimentally accessible scenario for testing and comparing a variety of impact models. Here we have presented one model that captures the relevant dynamics required for describing the motion of a real world billiard for arbitrary boundaries.  The model considers the more realistic situation of an inelastic, rotating, gravitational billiard in which there are retarding forces due to air resistance and friction.  We have used the model to investigate driven parabolic, wedge and hyperbolic billiards, and demonstrated that the parabola has stable, periodic motion, while the wedge and hyperbola (at high driving frequencies) appear chaotic.  The hyperbola, at low driving frequencies, has regular, periodic motion, and behaved similarly to the parabola.  The simple representation of the coefficient of restitution employed in the model resulted in good agreement with the recent experimental data of Feldt and Olafsen for all boundary shapes investigated \cite{Olafsen:1}, but not for secondary quantities derived from the data. The model also predicted additional collisions not detected by the data. Gorski and Srokowski \cite{Gorski:1} have also modeled the Feldt and Olafsen experiments. They employed a different model that included collisional energy loss, but ignored rotation. To achieve energy balance over long times and obtain qualitative agreement with the experimental data, in their work it was necessary to invoke an unrealistic energy dependece of the coeficient of restitution. Perhaps this was due to their unconventional model which applied the reduction in speed to the total velocity at collision, instead of the normal component (see Equation~(\ref{400:eq}) above). The assignment of the value of the coefficient of restitution introduces the most uncertainty in modeling the billiard-boundary system, and resolution of this problem will require additional experiments. It is interesting that the optimum numerical value is very different if rotation induced friction is not included. We will pursue this surprising effect in a future work.

\section{Acknowledgments}

This research was supported by a grant from Frank Ziglar, Jr. of North Carolina State University.  The authors would like to thank Dr. Jeff Olafsen of Baylor University for his helpful interactions.

\section{Appendix}

Here we list the equations of motion for impacts of billiards with moving boundaries.  The explanation that follows is a modified version of the impact theory set forth by Kane and Levinson for collisions between a sphere and stationary boundaries \cite{Kane:1}.  The theory considers the general three-dimensional case, but may be applied to two-dimensional systems by prescribing appropriate initial conditions for position and velocity.  Since the numerical simulations presented in this paper are two-dimensional, we will point out simplifications when appropriate.

\subsection{Background}

\begin{figure}[!hbp]
  \begin{center}
    \leavevmode
      \includegraphics[width=2.75in, angle=270]{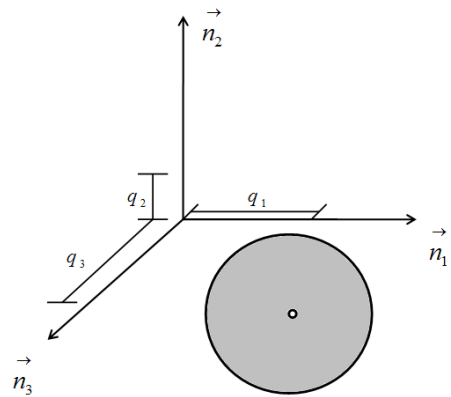}
    \caption{The inertial (or laboratory) coordinate frame.}
    \label{fig:unit_vectors1}
  \end{center}
\end{figure}

\begin{figure}[!hbp]
  \begin{center}
    \leavevmode
      \includegraphics[width=3.55in]{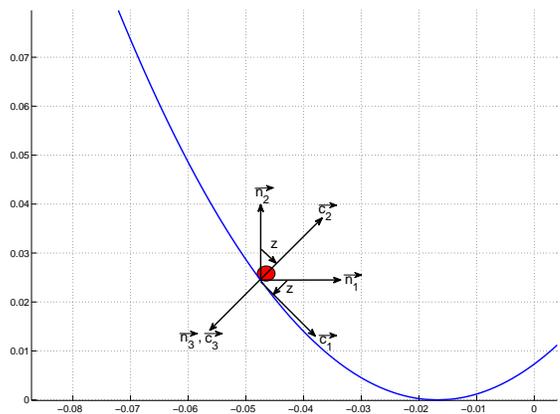}
    \caption{The coordinate frame at the collision point is denoted by the $\mathbf{c}$$-$frame.  Reference frames $\mathbf{c}$ and $\mathbf{n}$ are related to each other by angle $z$.}
    \label{fig:n_frame2}
  \end{center}
\end{figure}

Consider a sphere of mass m and radius b, whose motion is confined by a moving boundary.  The sphere has six degrees of freedom, three angles defining its orientation and three components defining its position.  The boundary is infinitely massive and its shape is arbitrary.  For the general billiard-boundary system, the inertial (or laboratory) frame is defined by three mutually perpendicular unit vectors ($\mathbf{n_{1}},\mathbf{n_{2}},\mathbf{n_{3}}$), where $\mathbf{n_{2}}$ is perpendicular to the plane formed by vectors $\mathbf{n_{1}}$ and $\mathbf{n_{3}}$, see Figure~\ref{fig:unit_vectors1}.  The reference frame at the collision point between the sphere and boundary is defined by the $\mathbf{c}$$-$frame, and is related to the inertial coordinate frame by a translation and rotation of coordinates.  It is more convenient to define the collision response in a frame moving with the boundary ($\mathbf{c}$-frame), oriented such that two of the unit vectors are locally parallel and orthogonal to the boundary surface at the collision point.  Figure~\ref{fig:n_frame2} shows the $\mathbf{c}$ and $\mathbf{n}$$-$frames for the two-dimensional case, where the frames are related by angle $z$.  For curved boundaries angle $z$ varies along the curve, and is uniquely determined for each collision.  The sphere's angular and translational velocities at the collision point may be expressed in terms of the generalized speeds
$u_{1}$,...,$u_{6}$, respectively, as
\begin{equation}
\label{1:function}
\boldsymbol{\omega}=u_{1}\mathbf{c_{1}}+u_{2}\mathbf{c_{2}}+u_{3}\mathbf{c_{3}}
\end {equation}
and
\begin{equation}
\label{2:function}
\mathbf{\\v}=u_{4}\mathbf{c_{1}}+u_{5}\mathbf{c_{2}}+u_{6}\mathbf{c_{3}}
\end {equation}
where $\mathbf{v}$ denotes the velocity of the sphere's center of mass.  The boundary's velocity at the collision point is given by
\begin{equation}
\label{4444:eq}
\mathbf{v^{'}}=u^{'}_{4}\mathbf{c_{1}}+u^{'}_{5}\mathbf{c_{2}}+u^{'}_{6}\mathbf{c_{3}}
\end {equation}
where (for our simulations) $u^{'}_{4}$, $u^{'}_{5}$ $u^{'}_{6}$ are found by taking components of $\mathbf{v^{'}} =$  $\omega Acos(\omega t$) $\mathbf{n_{1}}$.
Note that for the two-dimensional case, the following generalized speeds are zero: $u_{1}$, $u_{6}$ and $u^{'}_{6}$.

From Equations~(\ref{1:function}) and~(\ref{2:function}), we define the velocity of the point P of the sphere that comes into contact with the boundary as
\begin{eqnarray}
\label{3:function}
& \mathbf{v^{P}}&=\mathbf{v}+\boldsymbol{{\omega}}\times \boldsymbol{\rho} \nonumber\\
&& =(u_{4}\mathbf{c_{1}}+u_{5}\mathbf{c_{2}}+u_{6}\mathbf{c_{3}})  \nonumber\\
&& + (u_{1}\mathbf{c_{1}}+u_{2}\mathbf{c_{2}}+u_{3}\mathbf{c_{3}})\times-b\mathbf{c_{2}}  \nonumber\\
&& =(u_{4}+bu_{3})\mathbf{c_{1}}+u_{5}\mathbf{c_{2}}+(u_{6}-bu_{1})\mathbf{c_{3}}
\end{eqnarray}
Rewriting the result of Equation~(\ref{3:function}) in terms of the generalized speeds, we have
\begin{equation}
\label{31224:function898}
\mathbf{v^{P}}=(-b\mathbf{c_{3}})u_{1} + (0)u_{2} + (b\mathbf{c_{1}})u_{3} + (\mathbf{c_{1}})u_{4} + (\mathbf{c_{2}})u_{5} + (\mathbf{c_{3}})u_{6}
\end {equation}
Then from Equation~(\ref{31224:function898}), the partial velocities of P, labeled as $\mathbf{v^{P}_{r}}$ $(r=1,...,6)$, are determined by inspection and are simply the coefficients for each generalized speed.  For this problem $\mathbf{v^{P}_{1}}=-b\mathbf{c_{3}}$, $\mathbf{v^{P}_{2}}=0$, $\mathbf{v^{P}_{3}}=b\mathbf{c_{1}}$, $\mathbf{v^{P}_{4}}=\mathbf{c_{1}}$, $\mathbf{v^{P}_{5}}=\mathbf{c_{2}}$, $\mathbf{v^{P}_{6}}=\mathbf{c_{3}}$.
\\
\indent
Kane's impact model enables us to investigate the collision of a sphere as it impacts a boundary beginning at time $t_{1}$ and ending at time $t_{2}$.  Two dynamical equations essential to the model are the generalized impulse $I_{r}$ and the generalized momentum $p_{r}$.  The generalized impulse is generally applied to systems that are subjected to large action forces over a short time interval, and is defined as
\begin{equation}
\label{10:function}
\ I_{r}=\mathbf{v^{P}_{r}(t_{1})}\cdot\int^{t2}_{t1}\mathbf{R}\ dt   \ \ \ \ \ \ \ \    (r=1,...,6)
\end{equation}
where $\mathbf{v^{P}_{r}(t_{1})}$ is the partial velocity of the sphere at the point of contact with the boundary at time $t_{1}$, and $\int^{t2}_{t1}\mathbf{R}\ dt$ is the contact force exerted on the sphere by the boundary at their contact point
during the time interval [$t_{1}$, $t_{2}$].  Moreover, if we let $S_{i}$ = $\mathbf{c_{i}}\cdot\int^{t_{2}}_{t_{1}}\mathbf{R}\ dt$ (i=1,2,3), then we can define $\int^{t2}_{t1}\mathbf{R}\ dt$ as
\begin{equation}
\label{11:function}
\ \int^{t2}_{t1}\mathbf{R}\ dt=S_{1}\mathbf{c_{1}}+S_{2}\mathbf{c_{2}}+S_{3}\mathbf{c_{3}}\
\end{equation}
The generalized momentum is defined as follows:
\begin{equation}
\label{12:function}
\ p_{r}=\frac{\partial K}{\partial u_{r}}  \ \ \ \ \ \ \ \   (r=1,...,6)
\end{equation}
where K is the kinetic energy of the sphere and $u_{r}$ are the generalized speeds.  Integrating Equation~(\ref{10:function}) results in the following approximation connecting the generalized impulse to the generalized momentum:
\begin{equation}
\label{13:function}
\ I_{r}\approx p_{r}(t_{2})-p_{r}(t_{1}) \ \ \ \ \ \ \ \    (r=1,...,6)
\end{equation}
The approximation symbol in Equation~(\ref{13:function}) appears because forces that remain constant during the time interval [$t_{1}$, $t_{2}$] are regarded as negligible.
\\
\indent
In order to capture the sphere's motion at time $t_{2}$, two assumptions supplement the use of Equation~(\ref{13:function})
together with a complete description of the sphere's motion at time $t_{1}$.  The first assumption is the normal components of the velocity of approach $\mathbf{v_{A}}$ and separation $\mathbf{v_{S}}$ of the sphere, with respect to the boundary, have opposite directions, where the magnitudes are related by the following equation:
\begin{equation}
\label{14:function}
\ \mathbf{c_{2}}\cdot \mathbf{v_{S}}=-e\mathbf{c_{2}}\cdot \mathbf{v_{A}}
\end{equation}
In equation~(\ref{14:function}), e is the coefficient of restitution.  The second assumption determines if the sphere encounters no slipping or slipping at the point of contact with the boundary.  If there is no slipping at $t_{2}$, the following inequality must be satisfied:
\begin{equation}
\label{15:function}
\ \left|\boldsymbol{\tau}\right|<\mu\left|\boldsymbol{\nu}\right|
\end{equation}
where $\boldsymbol{\tau}$=$S_{1}\mathbf{c_{1}}+S_{3}\mathbf{c_{3}}$ is the tangential impulse,
$\boldsymbol{\nu}$=$S_{2}\mathbf{c_{2}}$ is the normal impulse
and $\mu$ is the coefficient of static friction.  Consequently,
\begin{equation}
\label{16:function}
\ \mathbf{c_{2}}\times(\mathbf{v_{S}}\times\mathbf{c_{2}})=0
\end{equation}
The equation states the tangential component of the velocity of separation is zero.
If inequality~(\ref{15:function}) is violated, slipping occurs at $t_{2}$, and $\boldsymbol{\tau}$ is expressed as
\begin{equation}
\label{17:function}
\\\boldsymbol{\tau}=-\mu^{'}\left|\boldsymbol{\nu}\right|\frac{\mathbf{c_{2}}\times(\mathbf{v_{S}}\times\mathbf{c_{2}})}{\left|\mathbf{c_{2}}\times(\mathbf{v_{S}}\times\mathbf{c_{2}})\right|}
\end{equation}
where the constant $\mu^{'}$ is the coefficient of kinetic friction.
\\
\indent
The model requires that the physical parameters b, m, J, e, $\mu$ and $\mu^{'}$, and the generalized speeds at
time $t_{1}$ are known, where J is the principal moment of inertia.  Then the motion of the sphere at time $t_{2}$ is fully defined by invoking Equations~(\ref{13:function}), ~(\ref{14:function}), ~(\ref{15:function}), ~(\ref{16:function}) and~(\ref{17:function}).

\subsection{Connection Formulas}

The connection formulas define the relationship between the billiard's pre-impact and post-impact velocities for both the no-slip and slip cases for collisions on moving boundaries.  For the no-slip case, the connection formulas are
\begin{equation}
\label{32:horz}
 u_{2}(t_{2})\approx u_{2}(t_{1})
\end{equation}
\begin{equation}
\label{35:horz}
 u_{5}(t_{2}) =  -eu_{5}(t_{1}) + u^{'}_{5}(t_{1})\left[1+e\right]
\end{equation}
\begin{equation}
\label{48:horz}
u_{3}(t_{2})\approx \frac{Ju_{3}(t_{1}) + mb[u^{'}_{4}(t_{1})-u_{4}(t_{1})]}{mb^{2}+J}
\end{equation}
\begin{equation}
\label{49:horz}
u_{4}(t_{2})= u^{'}_{4}(t_{1})-bu_{3}(t_{2})
\end{equation}
\begin{equation}
\label{53:horz}
u_{1}(t_{2})\approx \frac{Ju_{1}(t_{1}) + mb[u_{6}(t_{1})-u^{'}_{6}(t_{1})]}{mb^{2}+J}
\end{equation}
\begin{equation}
\label{54:horz}
u_{6}(t_{2})= u^{'}_{6}(t_{1})+bu_{1}(t_{2})
\end{equation}
\begin{equation}
\label{42:horz}
(S_{1}^{2}+S_{3}^{2})^{1/2}<\mu\left|S_{2}\right|
\end{equation}
\begin{eqnarray}\label{29:horz}
& S_{1}&\approx m[u_{4}(t_{2})-u_{4}(t_{1})]
\end{eqnarray}
\begin{eqnarray}\label{30:horz}
& S_{2}&\approx m[u_{5}(t_{2})-u_{5}(t_{1})]
\end{eqnarray}
\begin{eqnarray}\label{31:horz}
& S_{3}&\approx m[u_{6}(t_{2})-u_{6}(t_{1})]
\end{eqnarray}
where $e$ is the coefficient of restitution, $J$ is the principal moment of inertia, $m$ is the mass, $b$ is the radius, $S_{i}$ are the impulses for $i=1,2,3$ and $\mu$ is the coefficient of static friction.  For no slipping successive use of Equations~(\ref{32:horz}), ~(\ref{35:horz}), ~(\ref{48:horz}), ~(\ref{49:horz}), ~(\ref{53:horz}) and~(\ref{54:horz}) result in a set of values for $u_{1}$,..., $u_{6}$ at time $t_{2}$.  These values are valid if and only if inequality~(\ref{42:horz}) is satisfied for values of $S_{1}$, $S_{2}$ and $S_{3}$ given by Equations~(\ref{29:horz}),~(\ref{30:horz}) and~(\ref{31:horz}).  For the two-dimensional case, $u_{1}$, $u_{6}$, $u^{'}_{6}$ and $S_{3}$ are equal to zero.
\\
\indent
Otherwise, if inequality~(\ref{42:horz}) is violated, then the sphere is slipping at time $t_{2}$, and the quantities $u_{1}(t_{2})$, $u_{3}(t_{2})$, $u_{4}(t_{2})$, $u_{6}(t_{2})$, $S_{1}$ and $S_{3}$ must be recalculated using the relationships listed below:
\begin{eqnarray}\label{63:horz}
& \alpha= u_{4}(t_{1})+bu_{3}(t_{1})-u^{'}_{4}(t_{1})
\end{eqnarray}
\begin{eqnarray}\label{64:horz}
& \gamma= u_{6}(t_{1})-bu_{1}(t_{1})-u^{'}_{6}(t_{1})
\end{eqnarray}
\begin{eqnarray}\label{65:horz}
& k=\frac{1}{m}+\frac{b^{2}}{J}
\end{eqnarray}
\begin{equation}
\label{1012:horz}
S_{1}\approx -\frac{\mu^{'}\alpha \left|S_{2}\right|}{\left|\alpha \right|[1+(\gamma / \alpha)^{2}]^{1/2}}
\end{equation}
\begin{equation}
\label{69:horz}
 S_{3}\approx \frac{\gamma}{\alpha}S_{1}
\end{equation}
\begin{eqnarray}\label{57:horz}
& u_{1}(t_{2})\approx u_{1}(t_{1})-{bS_{3}}/{J}
\end{eqnarray}
\begin{eqnarray}\label{58:horz}
& u_{3}(t_{2})\approx u_{3}(t_{1})+{bS_{1}}/{J}
\end{eqnarray}
\begin{eqnarray}\label{59:horz}
& u_{4}(t_{2})\approx u_{4}(t_{1})+{S_{1}}/{m}
\end{eqnarray}
\begin{eqnarray}\label{60:horz}
& u_{6}(t_{2})\approx u_{6}(t_{1})+{S_{3}}/{m}
\end{eqnarray}
where $\alpha$, $\gamma$ and $k$ are constants and $\mu^{'}$ is the coefficient of kinetic friction.  Note that $u_{2}(t_{2})$, $u_{5}(t_{2})$ and $S_{2}$ are given by Equations~(\ref{32:horz}),~(\ref{35:horz}) and~(\ref{30:horz}) respectively, regardless of whether or not the sphere experiences no slipping or slipping at time $t_{2}$.  For the two-dimensional case, $u_{1}$, $u_{6}$, $u^{'}_{6}$ and $S_{3}$ are equal to zero.

\end{document}